\documentclass{article}
\usepackage{subfig}
\usepackage{graphicx}
\usepackage{spconf,amsmath,graphicx}
\usepackage{cite}
\usepackage{amssymb,amsfonts}
\usepackage{textcomp}
\usepackage{color}
\usepackage{fancyhdr}

\newcommand\norm[1]{\left\lVert#1\right\rVert}

\def\0{{\mathbf 0}}
\def\1{{\mathbf 1}}
\def\b{{\mathbf b}}
\def\m{{\mathbf m}}
\def\n{{\mathbf n}}
\def\p{{\mathbf p}}
\def\q{{\mathbf q}}
\def\s{{\mathbf s}}
\def\v{{\mathbf v}}

\def\y{{\mathbf y}}

\def\A{{\mathbf A}}
\def\B{{\mathbf B}}
\def\C{{\mathbf C}}

\def\H{{\mathbf H}}
\def\I{{\mathbf I}}

\def\cE{{\mathcal E}}
\def\cG{{\mathcal G}}

\def\cV{{\mathcal V}}
   
 \title{3D Point Cloud Super-Resolution via Graph Total Variation on Surface Normals}
\begin{document}
\ninept

\name{Chinthaka Dinesh{\small $~^{\#1}$}, Gene Cheung{\small $~^{\star2}$}, Ivan V. Baji\'{c}{\small $~^{\#3}$}
\thanks{This work was supported in part by NSERC Grant RGPIN-2016-04590.}
}%
\address{$^{\#}$\,Simon Fraser University, Burnaby, BC, Canada; $^{\star}$\,Department of Electrical Engineering \\ \& Computer Science, York University, Toronto, Canada\\
\fontsize{9}{9}\selectfont\ttfamily\upshape
\fontsize{9}{9}\selectfont\ttfamily\upshape
$^{1}$\,hchintha@sfu.ca; $^{2}$\,genec@yorku.ca; $^{3}$\,ibajic@ensc.sfu.ca\\}

\maketitle

\thispagestyle{empty}
\renewcommand{\headrulewidth}{0.0pt}
\thispagestyle{fancy}
\lhead{}
\chead{Accepted at IEEE ICIP'19, Taipei, Taiwan, Sep. 2019.}
\rhead{}
\cfoot{Copyright \copyright{ } 2019 IEEE. The original publication is available for download at ieeexplore.ieee.org.}

\begin{abstract}
Point cloud is a collection of 3D coordinates that are discrete geometric samples of an object's 2D surfaces. 
Using a low-cost 3D scanner to acquire data means that point clouds are often in lower resolution than desired for rendering on high-resolution displays. 
Building on recent advances in graph signal processing, we design a local algorithm for 3D point cloud super-resolution (SR). 
First, we initialize new points at centroids of local triangles formed using the low-resolution point cloud, and connect all points using a $k$-nearest-neighbor graph.
Then, to establish a linear relationship between surface normals and 3D point coordinates, we perform bipartite graph approximation to divide all nodes into two disjoint sets, which are optimized alternately until convergence.
For each node set, to promote piecewise smooth (PWS) 2D surfaces, we design a graph total variation (GTV) objective for nearby surface normals, under the constraint that coordinates of the original points are preserved. 
We pursue an augmented Lagrangian approach to tackle the optimization, and solve the unconstrained equivalent using the alternating method of multipliers (ADMM). 
Extensive experiments show that our proposed point cloud SR algorithm outperforms competing schemes objectively and subjectively for a large variety of point clouds. 
\end{abstract}


\section{Introduction}
Point cloud, acquired directly by off-the-shelf 3D scanners like Microsoft Kinect or estimated indirectly via stereo-matching algorithms~\cite{ji2017}, is a recently popular 3D visual signal representation for free viewpoint image rendering, and is investigated in industrial standards like MPEG\footnote{https://mpeg.chiariglione.org/standards/mpeg-i/point-cloud-compression}. 
Unlike 3D meshes, a point cloud is an unstructured list of 3D coordinates, and low-level processing tasks like compression~\cite{zhao_2017, Garcia2018, Sandri2019} and denoising~\cite{dinesh2018fast, dinesh2018} are challenging. 
In this paper, we focus on the little-studied point cloud super-resolution (SR) problem: how to increase the density of the 3D points while preserving the geometry of the underlying 3D object structure?

Compared to the formidable body of literature for image SR~\cite{Zweig2003, papyan2016, zhang2018, kang2013}, point cloud SR has received relatively little attention.
In~\cite{kil2006}, resolution of a point cloud is increased by combining similar low-res point cloud patches via simple local shifting and aggregation. 
This method relies on the existence of a large number of similar point cloud regions, which may not be true in practice.  
In~\cite{alexa2003, guennebaud2007, oztireli2009}, surface interpolation is accomplished using the Moving Least Squares (MLS) method. 
However, it is observed that these methods often over-smooth due to MLS interpolation. 
As one solution, \cite{huang2013} proposed an edge-aware point upsampling method by first sampling away from object boundaries, and then progress towards object edges and corners.
However, \cite{huang2013} is ad-hoc in methodology and still suffers from over-smoothing, albeit to a lesser extent.
Recently a deep learning based method \cite{yu2018} learns multilevel features per point and then upsamples the set of points via a multi-branch convolution unit implicit in the feature space. 
However, a large training dataset with similar geometric characteristics as the target point cloud is required, which may not be practical.  

In this paper, leveraging on recent progress in graph signal processing (GSP) \cite{cheung18,ortega2018} and our previous work on point cloud denoising~\cite{dinesh2018fast, dinesh2018}, we propose a fast local point cloud SR algorithm based on \textit{graph total variation} (GTV). 
First, we initialize new points at centroids of local triangles formed via \textit{Delaunay triangulation}\footnote{https\://en.wikipedia.org/wiki/Delaunay\_triangulation} using the low-res point cloud.
We connect all points using a $k$-nearest-neighbor graph. 
Then, to establish a tractable linear relationship between surface normals and 3D point coordinates (similarly done in~\cite{dinesh2018fast}), we perform \textit{bipartite graph approximation} \cite{zeng2017} to divide all nodes into two disjoint sets, which are optimized alternately until convergence.
For each node set, in order to promote piecewise smoothness (PWS) in reconstructed 2D surfaces, we design a GTV objective for nearby surface normals, under the constraint that 3D coordinates of the original points are preserved. 
We pursue an \textit{augmented Lagrangian} approach~\cite{boyd2011} and solve the unconstrained equivalent using the \textit{alternating method of multipliers} (ADMM) and proximal gradient decent~\cite{boyd2011,parikh2014}. 
We demonstrate in extensive experiments that our proposed point cloud SR algorithm outperforms competing schemes for a large variety of 3D point clouds objectively and subjectively.



\section{Preliminaries}
\label{sec:pre}

\noindent\textbf{3D Point Cloud:} We define a point cloud as a set of (roughly uniform) discrete samples of 3D coordinates on an object's 2D surface in 3D space. Denote by $\p= \left[ \p_{1}^{\top} \; \hdots \; \mathbf{p}_{N}^{\top} \right]^{\top} \in \mathbb{R}^{3N}$ the position vector for a full-resolution point cloud, where $\p_{i}\in\mathbb{R}^{3}$ is the 3D coordinate of a point $i$, and $N$ is the number of points in the point cloud. 
Similarly, denote by $\q= \left[ \mathbf{q}_{1}^{\top} \; \hdots \; \q_{M}^{\top} \right]^{\top} \in \mathbb{R}^{3M}$ the position vector for a low-res point cloud to the above one, where $\q_{i}\in\mathbb{R}^{3}$, and $M$, $M<N$, is the number of points in the low-res point cloud. We assume that the observed low-res $\q$ is noiseless. 
Full-resolution $\p$ and low-res $\q$ are related through the \textit{sampling matrix} $\mathbf{C}\in \{0, 1\}^{3M\times 3N}$; \textit{i.e.}, 
\begin{align}
\mathbf{Cp} = \mathbf{q}. 
\label{eq:SammplingM}
\end{align}

\noindent\textbf{Graph Construction from a 3D Point Cloud:} 
We construct a $k$-nearest-neighbor ($k$-NN) graph $\cG$ to connect 3D points in a point cloud \cite{wang2011}, so that each point can be filtered with neighboring points under a graph-structured data kernel \cite{cheung18,ortega2018}.
Specifically, consider an undirected graph $\mathcal{G}=(\cV, \cE)$ with node set $\mathcal{V}$ and edge set $\mathcal{E}$.
$\cG$ is specified by $(i,j,w_{i,j})$, where $i,j \in \cV$, $(i,j) \in \cE$, and $w_{i,j}\in\mathbb{R}^{+}$ is the weight of an edge that connects nodes $i$ and $j$. 
Each 3D point $i$ is represented by a node $i \in \cV$ and is connected via edges to its $k$ nearest neighbors $j$ in Euclidean distance, with weights $w_{i,j}$ that reflect inter-node similarities~\cite{dinesh2018}, \textit{i.e.},
\begin{equation}
w_{i,j}=\exp{\left\{-\frac{|\vert\mathbf{p}_{i}-\mathbf{p}_{j}|\vert_{2}^{2}}{\sigma_{p}^{2}}\right\}}\cos^{2}\theta_{i,j},
\label{eq:edge_weight}
\vspace{-5pt}
\end{equation}
where $\theta_{i,j}$ is the angle between surface normals $\mathbf{n}_i$ and $\mathbf{n}_j$ at respective nodes $i$ and $j$, and $\sigma_{p}$ is a parameter. 
In words, (\ref{eq:edge_weight}) states that weight $w_{i,j}$ is large (close to 1) if points $i$ and $j$ are physically close and associated normal vectors are similar, and small (close to 0) otherwise.
The weight definition in (\ref{eq:edge_weight}) is similar to bilateral filter weights defined in \cite{tomasi98} with domain and range filters.

\noindent\textbf{Surface Normals:} A surface normal $\n_i \in  \mathbb{R}^3$ of node $i$ is a vector that is perpendicular to the tangent plane at point $i$. 
Typically coordinates of the $k$ nearest neighbors of $i$ are used to compute $\n_i$ \cite{huang2001,kanatani2005,ouyang2005,gouraud1971,jin2005}. 
The most popular method is to fit a local plane to point $i$ and its $k$ nearest neighbors, and take the perpendicular vector to that plane (see \textit{e.g.} \cite{huang2001,kanatani2005,ouyang2005}). An alternative is to calculate the normal vector as the weighted average of the normal vectors of the triangles formed by point $i$ and different pairs of its neighbors (see \textit{e.g.} \cite{gouraud1971,jin2005}).

\section{Proposed Algorithm}
\label{sec:algo}

\subsection{Algorithm Overview}

We begin by adding a set of new interior points to the low-res point cloud $\mathbf{q}$ to populate the target full-res point cloud $\mathbf{p}$. 
Specifically, we construct a triangular mesh using Delaunay triangulation using points in $\q$, and then insert new points at the centroids of those triangles. 
Given an interpolated point cloud with appropriate density, we assume a graph signal smoothness prior to fine-tune the newly added 3D coordinates described as follows.  

Specifically, we assume that the underlying 2D surface of a 3D object is \textit{piecewise smooth} (PWS): 
surface normals $\n_i$ and $\n_j$ of two neighboring 3D points $i$ and $j$ should differ minimally over the 2D surface, except at the object boundaries (\textit{e.g.}, sides of a rectangular box), where the normal difference is large.
Hence, to promote small overall deviation with sparse but large changes, we design GTV to promote PWS\footnote{$l_1$ norm can better promote PWS than $l_2$ norm~\cite{rosset2007}.} of surface normals as follows (similarly done in~\cite{dinesh2018fast}):
\begin{equation}
|\vert\mathbf{n}|\vert_{\text{GTV}}=\sum_{i,j\in\mathcal{E}}w_{i,j}|\vert\mathbf{n}_{i}-\mathbf{n}_{j}|\vert_1,
\label{eq:main_TV}
\end{equation}

We then formulate our point cloud SR problem as a minimization of the defined GTV while enforcing the locations of observed points $\mathbf{q}$. 
Here, $\p$ is the optimization variable, and $\n_{i}$'s are functions of $\mathbf{p}$. 
Unfortunately, using state-of-art surface normal estimation methods, each $\n_{i}$ is a nonlinear function of $\p_{i}$ and its neighbors. 
Hence, it is difficult to formulate a clean convex optimization using the present form of GTV in (\ref{eq:main_TV}).

\begin{figure}[t]
\centering
\subfloat[original graph]{
\includegraphics[width=0.2\textwidth]{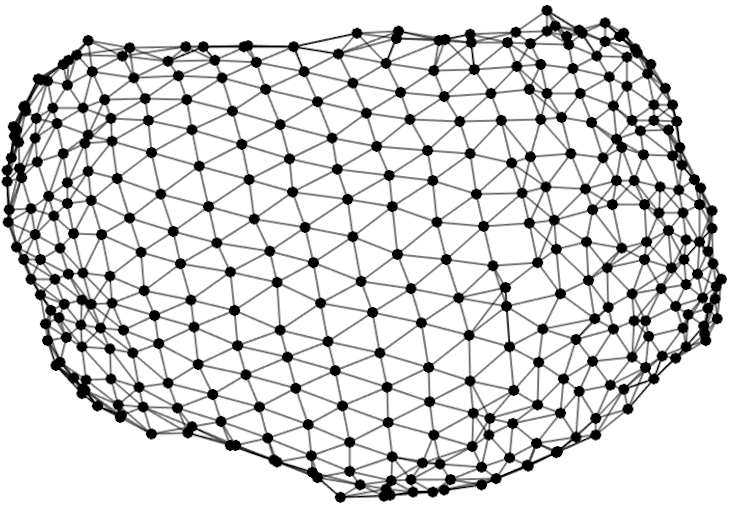}
\label{fig:pcd-day}}
\hspace{-10pt}
\subfloat[bipartite graph]{
\includegraphics[width=0.2\textwidth]{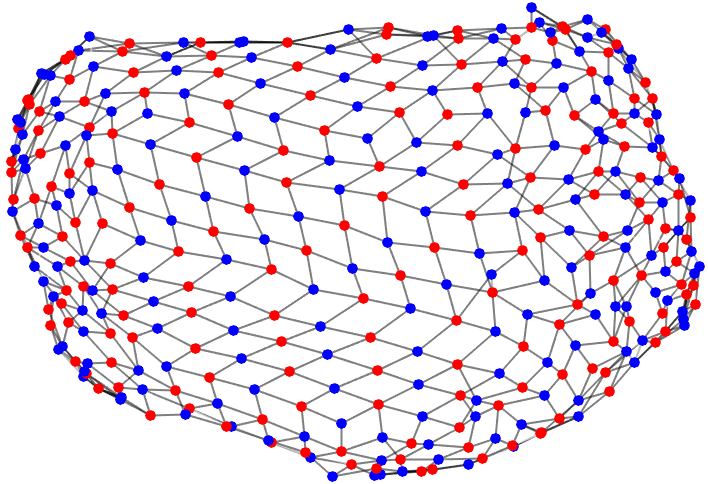}
\label{fig:pcd-on}}
\hspace{-6pt}
\caption[]{An example for bipartite graph approximation.} 
\label{fig:bipartite_graph}
\vspace{-12pt}
\end{figure}

To overcome this problem, following our previous work on point cloud denoising \cite{dinesh2018fast}, we first partition 3D points of the point cloud into two classes (say red and blue). 
When computing the normal for a red point, we employ as reference only neighboring blue points, and vice versa. 
Towards this goal, a $k$-neighborhood graph $\cG$ is initially constructed for all 3D points $\mathcal{V}$ in a point cloud. 
We next divide 3D points $\cV$ in a point cloud into two disjoint sets---red nodes $\cV_1$ and blue nodes $\cV_2$ where $\cV_1 \cup \cV_2 = \cV$ and $\cV_1 \cap \cV_2 = \emptyset$---via \textit{bipartite graph approximation} \cite{zeng2017}; \textit{i.e.}, finding a bipartite graph that best approximates the original non-bipartite graph in terms of Kullback-Leibler (KL) divergence\footnote{https://en.wikipedia.org/wiki/Kullback\%E2\%80\%93Leibler\_divergence}. 
An example for the bipartite graph approximation is shown in Fig.\;\ref{fig:bipartite_graph}. 
Doing so means that each red node $i \in \cV_1$ has in its neighborhood two or more blue nodes, using which node $i$'s surface normal $\n_i$ can be written as a linear function of its 3D coordinate $\p_i$, \textit{i.e.}, 
\begin{align}
\n_i = \A_i \p_i + \b_i,
\label{eq:normali}
\end{align}
where $\mathbf{A}_i\in \mathbb{R}^{3\times 3}$ and $\mathbf{b}_i\in \mathbb{R}^3$ are computed from coordinates of the neighbouring blue nodes. 
See \cite{dinesh2018fast} for further details.

\subsection{Optimization Objective}

We next construct a \textit{new} $k$-NN graph $\mathcal{G}_1~=~(\mathcal{V}_{1},\mathcal{E}_{1})$ connecting only neighboring red nodes, where $\mathcal{E}_{1}$ is the set of edges of the new graph. 
Edge weights $w_{i,j}$ are computed using (\ref{eq:edge_weight}), where $\theta_{i,j}$ is computed using initial surface normal estimates. 
Denote by $\m_{i,j}$ the difference between normals $\mathbf{n}_i$ and $\mathbf{n}_j$ of two connected nodes $i$ and $j$ in $\mathcal{G}_1$, \textit{i.e.}, $\m_{i,j} = \n_i - \n_j$, where $(i,j)\in \mathcal{E}_1$.
Given that each normal $\n_i$ is a linear function of node $i$'s 3D coordinates $\p_i$ (\ref{eq:normali}),  we can write $\m = \left[ \{\m_{i,j}\} \right]^{\top}$ also as a linear function of $\p = \left[ \{\p_i \} \right]^{\top}$:
\begin{align}
\m = \B \p + \v,
\label{eq:m}
\end{align}

Given the relationship between low-res $\q$ and full-res $\p$ through sampling matrix (\ref{eq:SammplingM}), we can summarize the two constraints compactly as a single equality constraint as follows:
\begin{align}
\underbrace{\left[\begin{array}{cc}
\I & -\B \\
0 & \C
\end{array}
\right]}_{\H}    
\underbrace{\left[\begin{array}{c}
\m \\
\p
\end{array}
\right]}_{\s} = 
\underbrace{\left[\begin{array}{c}
\v \\
\q
\end{array}
\right]}_{\mathbf{d}}
\label{eq:constraints}
\end{align}
where $\s = [\m \;\; \p]^{\top}$ is the combined optimization variable. 

Our objective is to minimize the GTV of surface normals:
\begin{equation}
\begin{split}
&\quad \min_{\s} \sum_{i,j} w_{i,j} \| \m_{i,j} \|_1
\label{eq:SR_lost}
\end{split}
\end{equation}
subject to linear constraint (\ref{eq:constraints}). 
In our optimization framework, we first solve (\ref{eq:SR_lost}) for red nodes while the positions of blue nodes are fixed. 
Then, using the newly solved red nodes' positions, we initialize the normals for blue nodes and construct another $k$-NN graph for blue nodes. 
Then (\ref{eq:SR_lost}) is solved for blue nodes while the positions of red nodes fixed. 
The two node sets are alternately optimized until convergence. 
We next discuss how to solve (\ref{eq:SR_lost}) for one node set.

\subsection{Optimization Algorithm}

To solve the problem in~(\ref{eq:SR_lost}), we pursue an augmented Lagrangian approach and rewrite the constrained problem (\ref{eq:SR_lost}) into the following unconstrained form:
\begin{equation}
\min_{\s} \sum_{i,j} w_{i,j} \| \m_{i,j} \|_1+\mathbf{y}^{\top}(\mathbf{Hs-d})+\frac{\rho}{2}\norm{\mathbf{Hs-d}}_2^2, 
\label{eq:cost_AL}
\end{equation}
where $\y$ is a dual variable and $\rho$ is a Lagrange multiplier. 
The unconstrained version (\ref{eq:cost_AL}) can now be solved using ADMM and proximal gradient decent.  
According to~(\ref{eq:constraints}), we can rewrite $\norm{\mathbf{Hs-d}}_2^2$ as
\begin{equation}
\norm{\mathbf{Hs-d}}_2^2=\norm{\mathbf{Bp+v-m}}_2^2+\norm{\mathbf{Cp-q}}_2^2.   
\end{equation}
Similarly, according to~(\ref{eq:constraints}), $\mathbf{y}^{\top}(\mathbf{Hs-d})$ can be rewritten as
\begin{equation}
 \mathbf{y}^{\top}(\mathbf{Hs-d})=\mathbf{y}_{1}^{\top}(\mathbf{Bp+v-m})+\mathbf{y}_2^{\top}(\mathbf{Cp-q}),
\end{equation}
where $\mathbf{y}^{\top}=[\mathbf{y}_1^{\top}\hspace{8pt} \mathbf{y}_2^{\top}]$. We can thus rewrite (\ref{eq:cost_AL}) as
\begin{equation}
\begin{split}
\min_{\m, \p} \sum_{i,j} w_{i,j} \| \m_{i,j} \|_1+\mathbf{y}_{1}^{\top}(\mathbf{Bp+v-m})+\\ \mathbf{y}_2^{\top}(\mathbf{Cp-q})+\frac{\rho}{2}\norm{\mathbf{Bp+v-m}}_2^2+\frac{\rho}{2}\norm{\mathbf{Cp-q}}_2^2.
\end{split}
\label{eq:SR_AL_solve}
\end{equation}

As typically done in ADMM approaches, we solve (\ref{eq:SR_AL_solve}) by alternately minimizing $\mathbf{p}$ and $\mathbf{m}$ and updating $\mathbf{y}$ one at a time in turn until convergence.\\
\textbf{$\mathbf{p}$ minimization}: To minimize $\mathbf{p}$ having $\mathbf{m}^{k}$ and $\mathbf{u}^{k}$ fixed, we take the derivative of (\ref{eq:SR_AL_solve}) with respect to $\mathbf{p}$, set it to $0$ and solve for the closed form solution $\mathbf{p}^{k+1}$:
\begin{equation}
\begin{split}
\rho(\mathbf{B}^{\top}\mathbf{B}+\mathbf{C}^{\top}\mathbf{C})\mathbf{p}^{k+1}=\mathbf{C}^{\top}(\rho\mathbf{q}-\mathbf{y}_2)+\\
\mathbf{B}^{\top}(\rho\mathbf{m}^{k}-\rho\mathbf{v}-\mathbf{y}_1)
\end{split}
\label{eq:p_mini}
\end{equation}
One can show that matrix $(\mathbf{B}^{\top}\mathbf{B}~+~\mathbf{C}^{\top}\mathbf{C})$ is positive definite (PD) but can be numerically unstable with a large condition number (the ratio of the largest eigenvalue to the smallest). 
To obtain a stable solution nonetheless for~(\ref{eq:p_mini}), we use the \textit{iterative refinement} method proposed in~\cite{parikh2014}.\\
\textbf{$\mathbf{m}$ minimization}: Keeping $\mathbf{p}^{k+1}$ and $\mathbf{y}^{k}$ fixed, the minimization of $\mathbf{m}$ becomes:
\begin{equation}
\begin{split}
\min_{\m} \mathbf{y}_{1}^{k^{\top}}(\mathbf{Bp^{k+1}+v-m})+\frac{\rho}{2}\norm{\mathbf{Bp^{k+1}+v-m}}^{2}_{2}+\\
\sum_{i,j} w_{i,j} \norm{\m_{i,j}}_{1}
\end{split}
\label{eq:SR_AL_m_min}
\end{equation}
where the first two terms are convex and differentiable, and the third term is convex but non-differentiable. We can thus use \textit{proximal gradient}\cite{parikh2014} to solve (\ref{eq:SR_AL_m_min}). The first two terms have gradient $\Delta_{\mathbf{m}}$:
\begin{equation}
\Delta_{\mathbf{m}}(\mathbf{p}^{k+1},\mathbf{m},\mathbf{y}_1^{k})=-\rho(\mathbf{Bp}^{k+1}+\mathbf{v-m})-\mathbf{y}_{1}^{k}.
\end{equation}

We can now define a proximal mapping $\text{prox}_{g,t}(\mathbf{m})$ for a convex, non-differentiable function $g()$ with step size $t$ as:
\begin{equation}
\text{prox}_{g,t}(\mathbf{m})=\text{arg}\min_{\theta}\left\{g(\theta)+\frac{1}{t}|\vert\theta-\mathbf{m}\vert|_{2}^{2}\right\}
\end{equation}
We know that for our weighted $l_{1}$-norm in (\ref{eq:SR_AL_m_min}), the proximal mapping is just a soft thresholding function:
\begin{equation}
\text{prox}_{g,t}(m_{i,j,r})=\begin{cases}
m_{i,j,r}-t w_{i,j} &\text{if} \hspace{5pt} m_{i,j,r}>t w_{i,j}\\
0 &\text{if} \hspace{5pt} \vert m_{i,j,r}\vert\leq t w_{i,j}\\
m_{i,j,r}+t w_{i,j} &\text{if} \hspace{5pt} m_{i,j,r}<-t w_{i,j},
\end{cases}
\end{equation}
where $m_{i,j,r}$ is the $r$-th entry of $\mathbf{m}_{i,j}$. We can now update $\mathbf{m}^{k+1}$ as:
\begin{equation}
\mathbf{m}^{k+1}=\text{prox}_{g,t}(\mathbf{m}^{k}-t\Delta_{\mathbf{m}}(\mathbf{p}^{k+1},\mathbf{m}^{k},\mathbf{y}_1^{k})).
\label{eq:m_update}
\end{equation}
We compute (\ref{eq:m_update}) iteratively until convergence.\\
\textbf{$\mathbf{y}$-update}: Finally, we can update $\mathbf{y}^{k+1}$ simply:
\begin{equation}
\mathbf{y}^{k+1}=\mathbf{y}^{k}+\rho(\mathbf{As}^{k+1}-b).
\label{eq:u_update}
\end{equation}
$\mathbf{p}$, $\mathbf{m}$ and $\mathbf{y}$ are iteratively optimized in turn using (\ref{eq:p_mini}), (\ref{eq:m_update}) and (\ref{eq:u_update}) until convergence.

\section{Experimental Results}

\begin{table}[t]
\centering
\caption{C2C ($\times 10^{-1}$) of different models}
\label{my-label}
\begin{tabular}{c|c|c|c|c|c}
\hline
Model    & Low & APSS           & RIMLS & Initial & Prop.       \\ \hline
Bunny    & 1.47 & 1.31          & 1.22 & 1.27        & \textbf{1.14}    \\ \hline
Dragon & 1.52 & 1.43          & 1.34 & 1.45        & \textbf{1.25}    \\ \hline
Armadillo       & 1.49 & 1.38 & 1.30 & 1.37        & \textbf{1.21}          \\ \hline
Buddha & 1.55 & 1.42          & 1.32  & 1.46       & \textbf{1.27} \\ \hline
Asian Drag.   & 1.56 & 1.41          & 1.39 & 1.44        & \textbf{1.31}    \\ \hline
Lucy & 1.58 & 1.46          & 1.40 & 1.51       & \textbf{1.28}    \\ \hline
\end{tabular}
\label{tab:C2C}
\vspace{-15pt}
\end{table}

\begin{table}[t]
\centering
\caption{C2P ($\times 10^{-2}$) of different models}
\label{my-label}
\begin{tabular}{c|c|c|c|c|c}
\hline
Model    & Low & APSS           & RIMLS & Initial & Prop.       \\ \hline
Bunny    & 3.92 & 2.40          & 2.28 & 2.49        & \textbf{1.87}    \\ \hline
Dragon & 2.81 & 1.71          & 1.58 & 1.77        & \textbf{1.26}    \\ \hline
Armadillo       & 2.55 & 1.42 & 1.30 & 1.37        & \textbf{1.07}          \\ \hline
Buddha & 3.11 & 1.93          & 1.81  & 1.88       & \textbf{1.38} \\ \hline
Asian Drag.   & 2.76 & 1.62          & 1.47 & 1.57        & \textbf{1.31}    \\ \hline
Lucy & 3.53 & 2.36          & 2.11 & 2.44       & \textbf{1.68}    \\ \hline
\end{tabular}
\label{tab:C2P}
\vspace{-12pt}
\end{table}

\begin{figure*}[t]
\centering
\subfloat[ground truth]{
\includegraphics[width=0.19\textwidth]{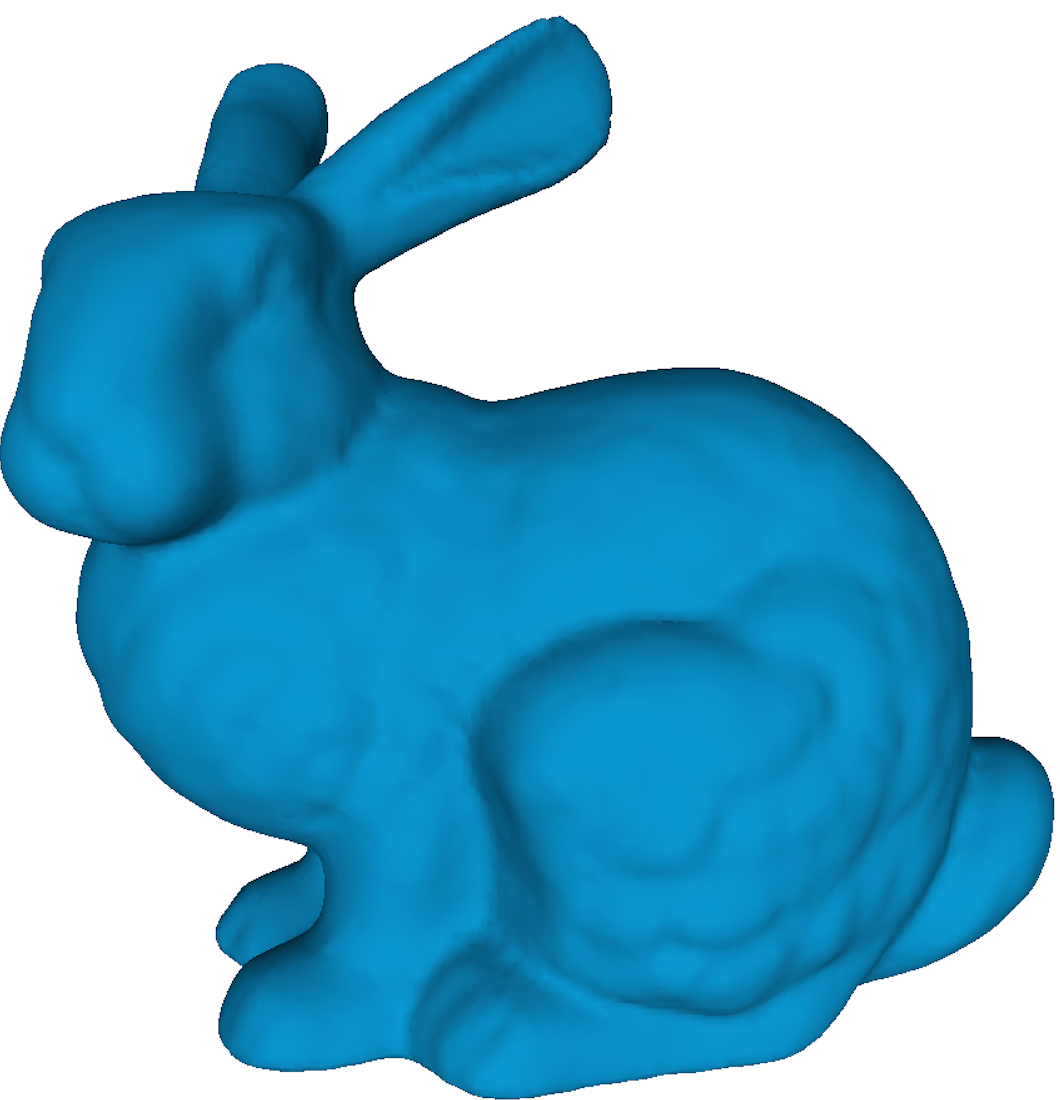}
\label{fig:pcd-day}}
\hspace{-7pt}
\subfloat[low resolution]{
\includegraphics[width=0.19\textwidth]{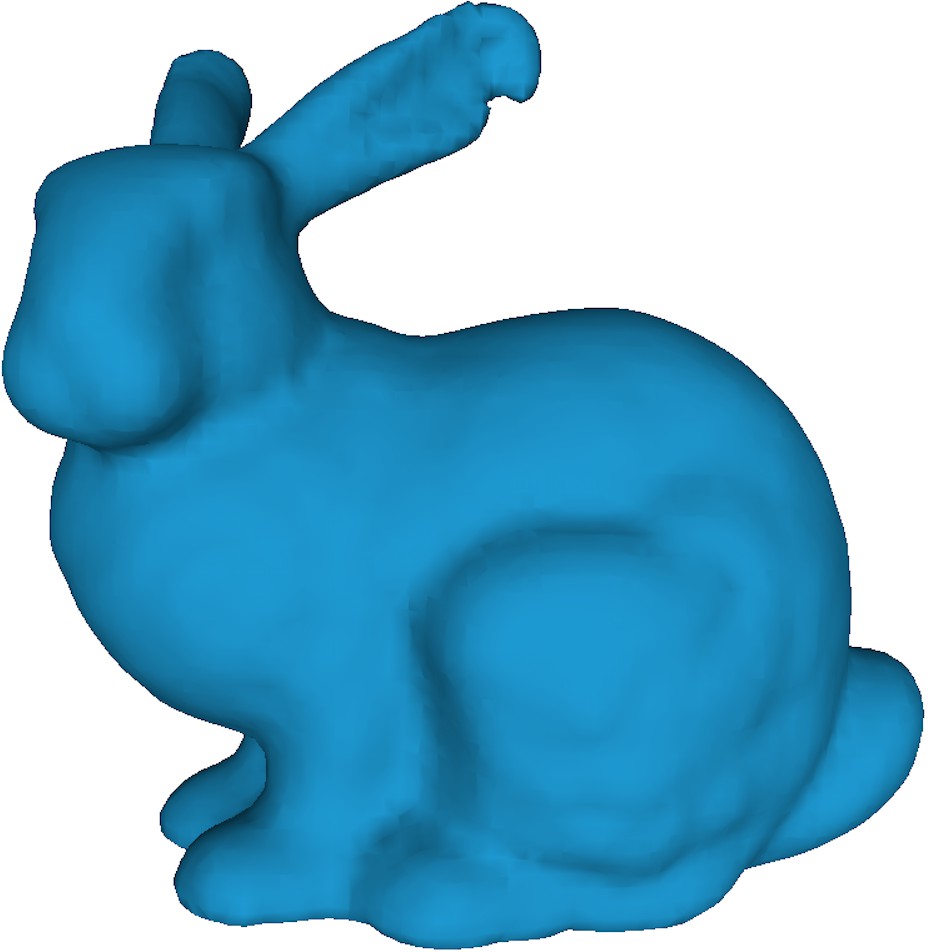}
\label{fig:pcd-on}}
\hspace{-7pt}
\subfloat[APSS]{
\includegraphics[width=0.19\textwidth]{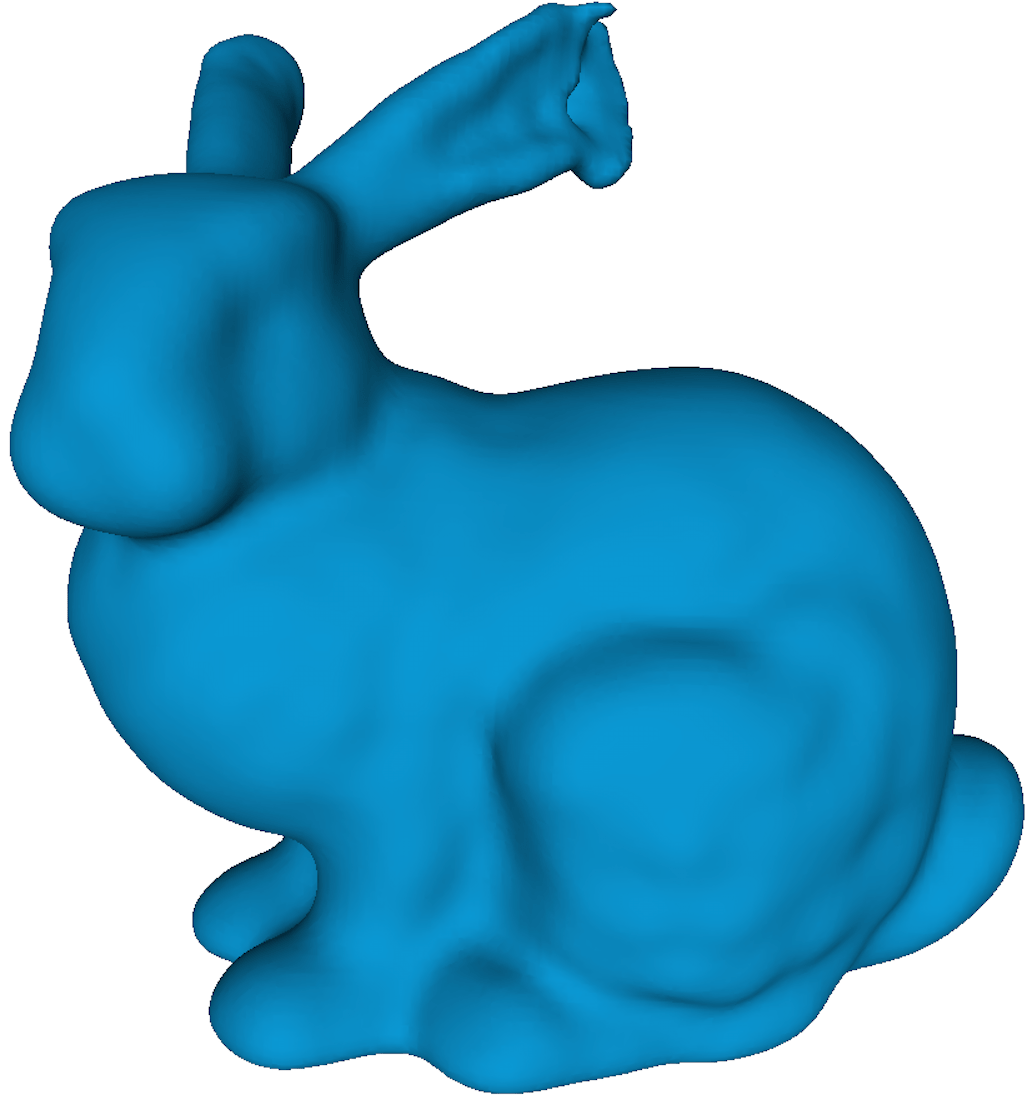}
\label{fig:pcd-day}}
\hspace{-7pt}
\subfloat[RIMLS]{
\includegraphics[width=0.19\textwidth]{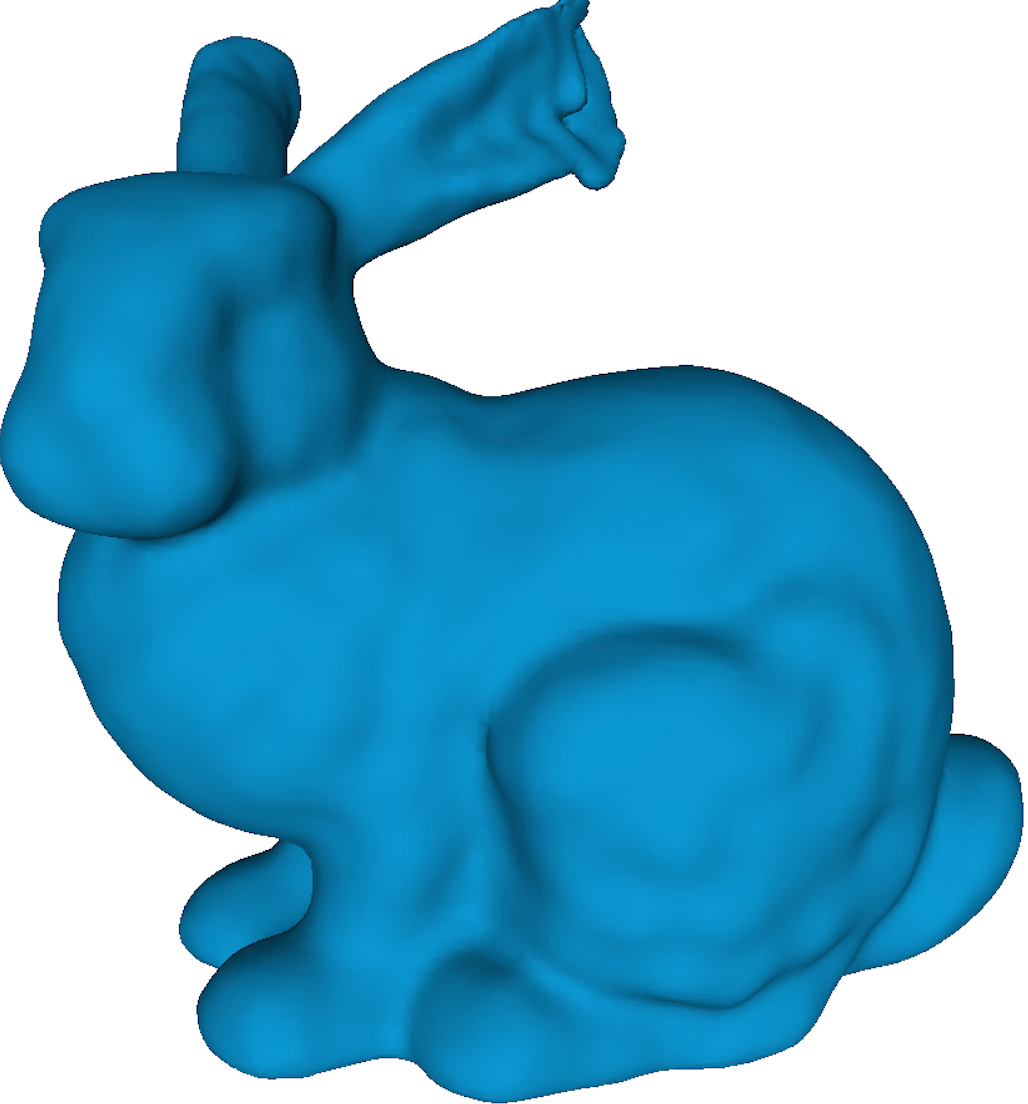}
\label{fig:pcd-on}}
\hspace{-7pt}
\subfloat[proposed]{
\includegraphics[width=0.19\textwidth]{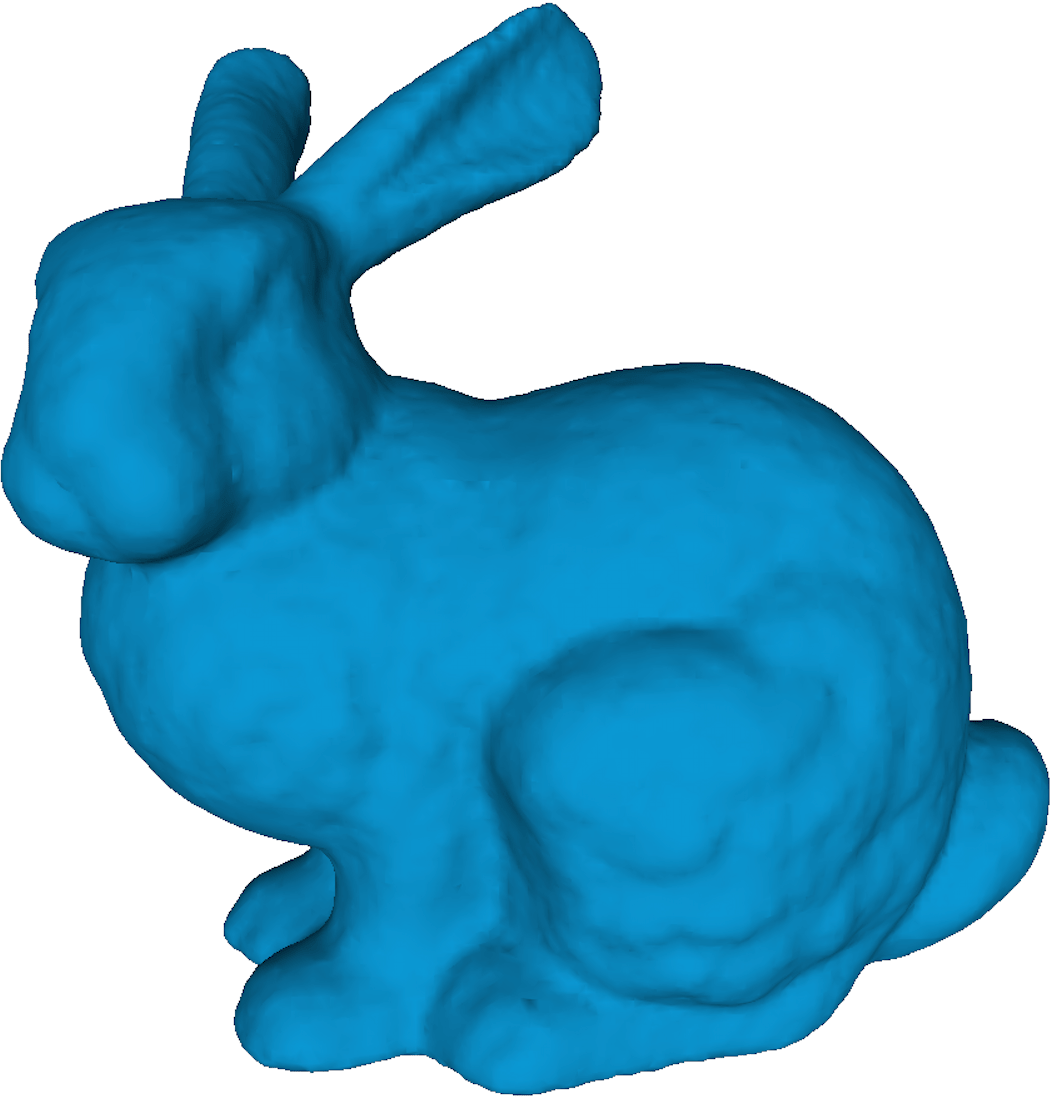}
\label{fig:tvlcd-day}}
\hspace{-9pt}
\vspace{-5pt}
\caption[Super resolution results illustration for Bunny model]{Super resolution results illustration for Bunny model; a surface is fitted over the point cloud for better visualization.} 
\label{fig:bunny}
\vspace{-0pt}
\end{figure*}


\begin{figure*}[t!]
\centering
\subfloat[ground truth]{
\includegraphics[width=0.23\textwidth]{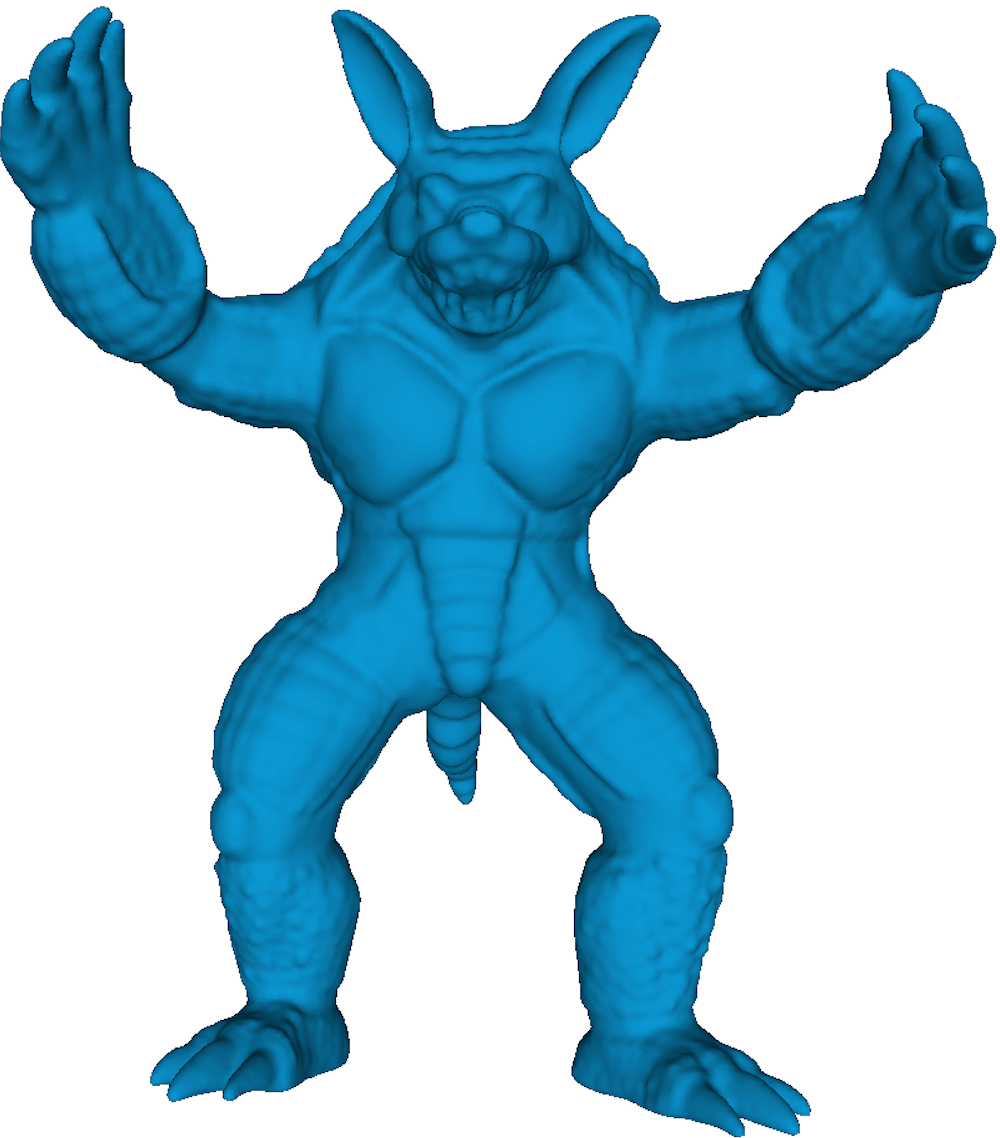}
\label{fig:pcd-day}}
\hspace{-0pt}
\subfloat[low resolution]{
\includegraphics[width=0.23\textwidth]{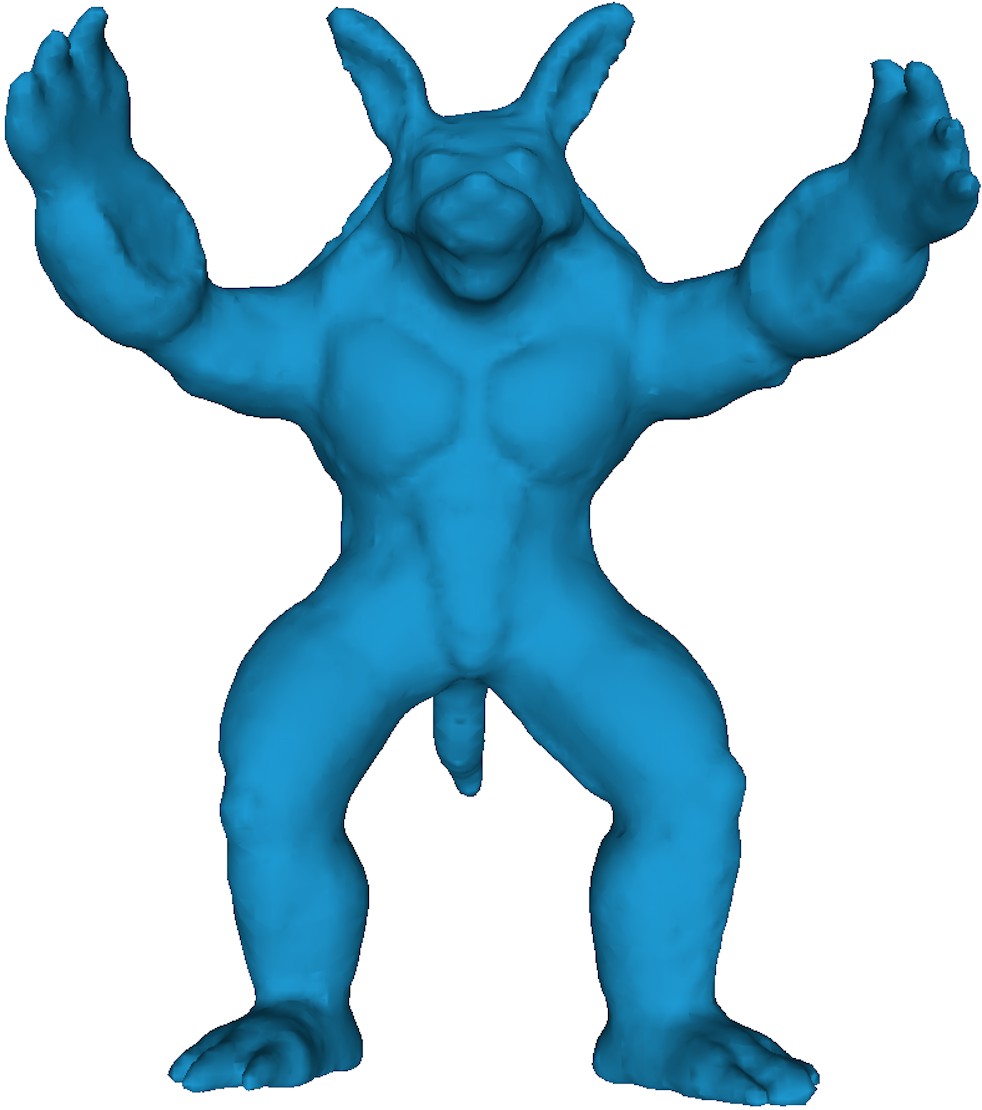}
\label{fig:pcd-on}}
\hspace{10pt}
\subfloat[APSS]{
\includegraphics[width=0.23\textwidth]{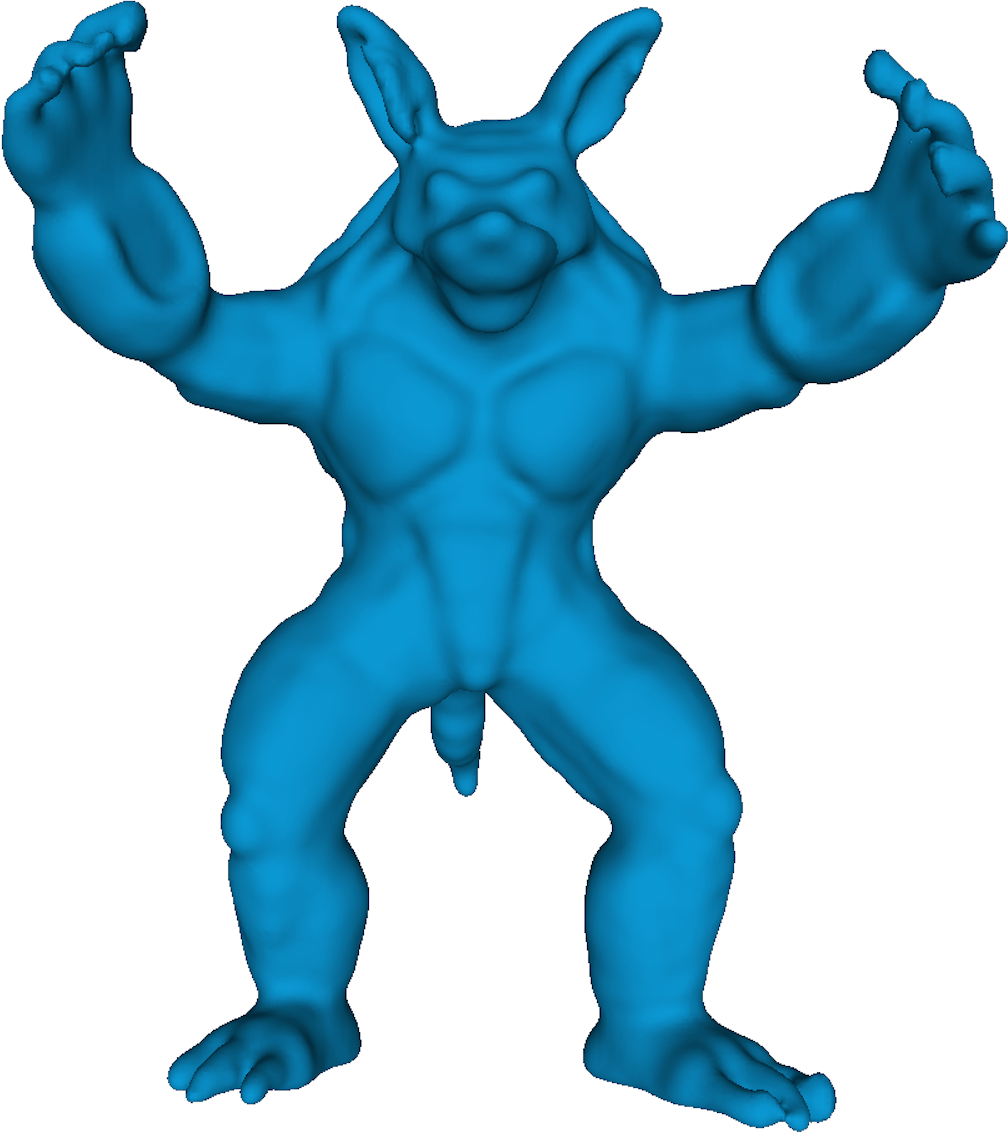}
\label{fig:pcd-day}}
\hspace{10pt}
\subfloat[RIMLS]{
\includegraphics[width=0.23\textwidth]{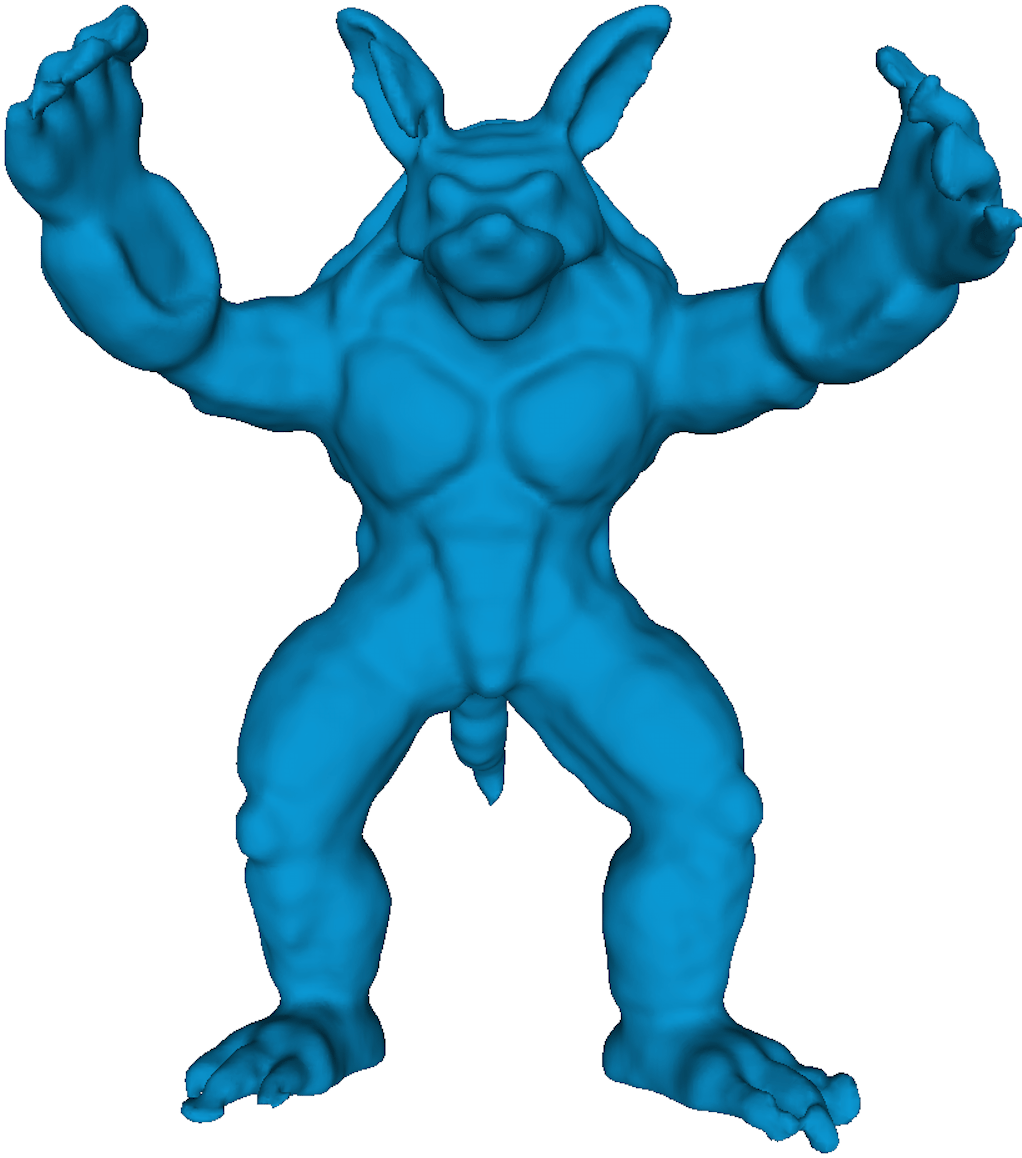}
\label{fig:pcd-on}}
\hspace{10pt}
\subfloat[proposed]{
\includegraphics[width=0.23\textwidth]{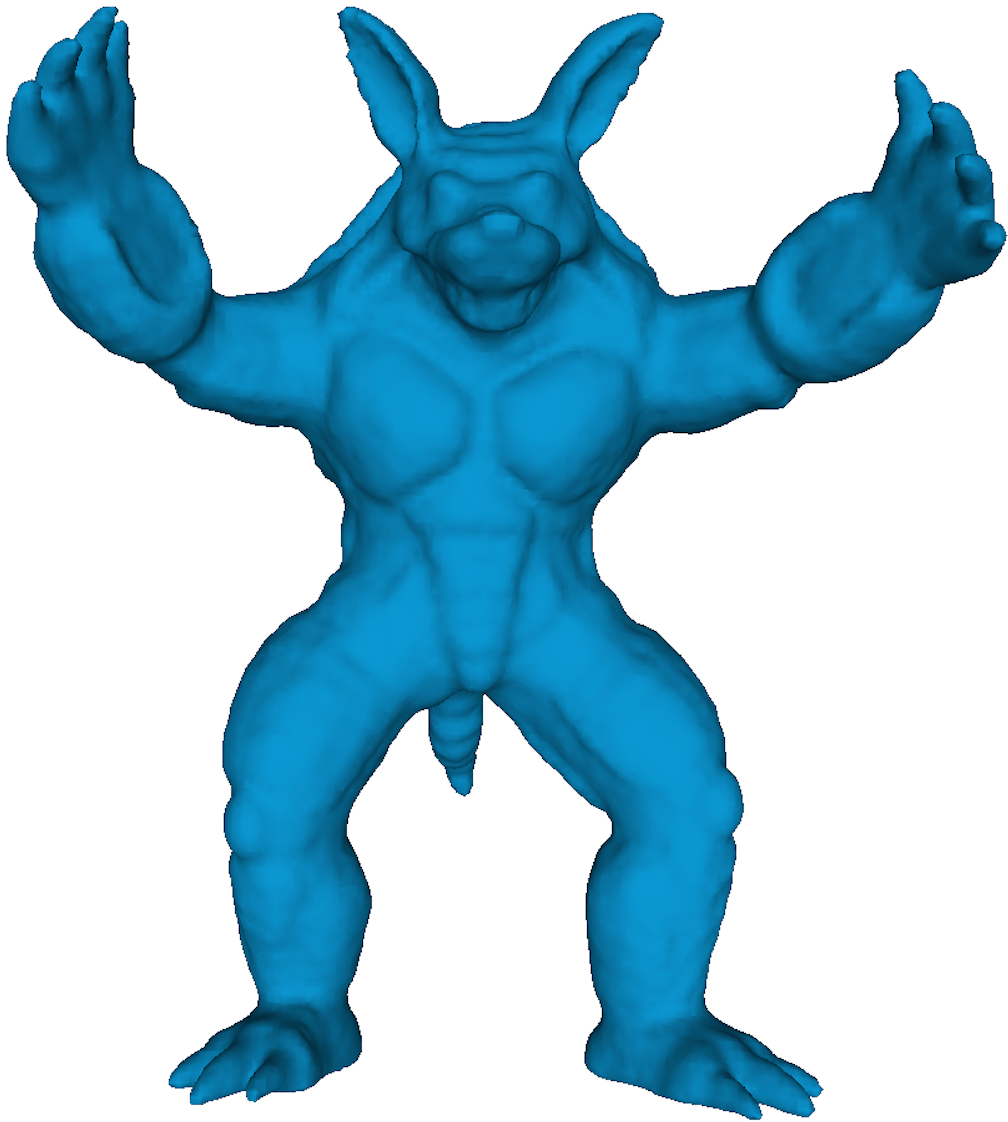}
\label{fig:tvlcd-day}}
\hspace{-0pt}
\vspace{-5pt}
\caption[Super resolution results illustration for Armadillo model]{Super resolution results illustration for Armadillo model; a surface is fitted over the point cloud for better visualization.} 
\label{fig:ar}
\vspace{-10pt}
\end{figure*}

\label{sec:results}

The computed SR  point clouds are compared against point cloud models obtained using competing schemes APSS~\cite{guennebaud2007} and RIMLS \cite{oztireli2009}, as well as initial point cloud (after adding new points to the centroids of triangles). Point cloud models we use are Bunny, Dragon, Armadillo, Happy Buddha, Asian Dragon, and Lucy, provided in~\cite{levoy2005}. 
All point cloud models are first rescaled, so that each tightly fits inside a bounding box with the same diagonal. 
Both quantitative and visual comparisons are presented.

For numerical comparisons, we measure the point-to-point (denoted as C2C) error and point to plane (denoted as C2P) error \cite{tian2017} between ground truth and SR results. 
In C2C error, we first measure the average of the squared Euclidean distances between ground truth points and their closest points in the SR cloud, and also that between the points of the SR cloud and their closest ground truth points. 
Then the larger value among these two measures is computed as C2C error. 
In C2P error, we first measure the average of the squared Euclidean distances between ground truth points and tangent planes at their closest points in the SR cloud, and also that between the points of the SR cloud and tangent planes at their closest ground truth points. 
Again, the larger value between these two measures is computed as the C2P error.

First, the ground truth point cloud are downsampled to 30\% of the original points using Poisson Disk Sampling method that was implemented in MeshLab software tool~\cite{cignoni2008}. 
Then, downsampled point clouds are upsampled to get the same number of points of their ground truths by using the proposed method and existing methods. 
Numerical results are shown in Table\;\ref{tab:C2C} (with C2C error) and Table\;\ref{tab:C2P} (with C2P error).
We observe that our proposed method has the lowest C2C and C2P errors. 
In each experiment, the selected parameters are $\rho=5$, $t=0.1$, and $k=8$ when constructing $k-$NN graphs.

Apart from the numerical comparison, visual results for Bunny and Armadillo models are shown in Fig.\;\ref{fig:bunny} and \ref{fig:ar}, respectively. We see that APSS and RIMLS schemes generate overly smooth models compared to our proposed method. 
Further, existing methods result in distorted surfaces with some details lost. 
This can be clearly seen at the left ear of bunny model and and the fingers (legs and hands) of armadillo model. 
On the other hand, using our proposed method, the details are well preserved without over-smoothing.

\section{Conclusion}
\label{sec:concl}

We pursue a graph-based approach to tackle the point cloud super-resolution (SR) problem, where the density of the point cloud is increased while preserving piecewise smoothness (PWS) of the intended object's 2D surface. 
Specifically, we first initialize new points at centroids of local triangles, and construct a $k$-nearest-neighbor graph to connect all 3D points. 
We then divide the nodes into two disjoint sets (red and blue) via bipartite graph approximation, which are subsequently optimized in an alternating manner.
Surface normal for each red node can now be defined as a linear function of its 3D coordinate using neighboring blue nodes as reference.
We can thus formulate a convex optimization problem, with a graph total variation (GTV) on surface normals as objective and a linear equality constraint to enforce original point coordinates.
Experimental results show that our proposed method outperforms competing schemes objectively and subjectively for a variety of 3D point clouds.

\bibliographystyle{IEEEtran}
\bibliography{refs}

\end{document}